\documentclass[aps,prl,twocolumn,showpacs,superscriptaddress,groupedaddress,floatfix]{revtex4}   %for PRL and submission

\pdfoutput=1 %required for arXiv submission?

\usepackage{bibentry}
\usepackage[utf8]{inputenc}
\usepackage[LGR, T1]{fontenc}
\usepackage[greek,english]{babel}
\usepackage{CJKutf8}
\usepackage{textcomp}
\usepackage{textgreek}

\usepackage{graphicx}  % needed for figures
\usepackage{dcolumn}   % needed for some tables
\usepackage{bm}        % for math
\usepackage{amssymb}   % for math
\usepackage{amsmath}

\usepackage[colorlinks = true,
            linkcolor = blue,
            urlcolor  = blue,
            citecolor = blue,
            anchorcolor = blue]{hyperref}

\usepackage{listings}
\usepackage{color}

\definecolor{myred}{rgb}{1,0.1,0}
\definecolor{myred2}{rgb}{1,0.1,0}

\begin{document}

% The following information is for internal review, please remove them for submission
%\widetext
%\leftline{Version 16 as of \today}
%\leftline{Primary authors: P. Allain, D. Damiron, H. Kawakatsu}

%\leftline{Comment to {\tt kawakatu@iis.u-tokyo.ac.jp} }

\title{Color Atomic Force Microscopy with on-the-fly Morse parameters mapping}
\author{P. E. Allain}
\author{D. Damiron}
\affiliation{LIMMS/CNRS UMI2820 Institute of Industrial Science, The University of Tokyo, 4-6-1 Komaba, Meguro-ku, Tokyo 153-8505, Japan}
\affiliation{CIRMM, Institute of Industrial Science, The University of Tokyo, 4-6-1 Komaba, Meguro-ku, Tokyo 153-8505, Japan}

\author{Y. Miyazaki }
\author{K. Kaminishi }
\affiliation{CIRMM, Institute of Industrial Science, The University of Tokyo, 4-6-1 Komaba, Meguro-ku, Tokyo 153-8505, Japan}

\author{F. V. Pop}
\affiliation{LIMMS/CNRS UMI2820 Institute of Industrial Science, The University of Tokyo, 4-6-1 Komaba, Meguro-ku, Tokyo 153-8505, Japan}
\affiliation{CIRMM, Institute of Industrial Science, The University of Tokyo, 4-6-1 Komaba, Meguro-ku, Tokyo 153-8505, Japan}

\author{D. Kobayashi}
\affiliation{CIRMM, Institute of Industrial Science, The University of Tokyo, 4-6-1 Komaba, Meguro-ku, Tokyo 153-8505, Japan}

\author{N. Sasaki}
\affiliation{Department of Engineering Science, the University of Electro-Communication, Chofu, Japan}

\author{H. Kawakatsu}
\affiliation{LIMMS/CNRS UMI2820 Institute of Industrial Science, The University of Tokyo, 4-6-1 Komaba, Meguro-ku, Tokyo 153-8505, Japan}
\affiliation{CIRMM, Institute of Industrial Science, The University of Tokyo, 4-6-1 Komaba, Meguro-ku, Tokyo 153-8505, Japan}

%\date{\today}

\begin{abstract}

Atomic Force Microscopy has enabled 2D imaging at the sub-molecular level, and 3D mapping of the potential field. However, fast identification of the surface still remains a challenging topic. In this paper, as a step towards implementation of such function, we introduce a control scheme and mathematical treatment of the acquired data that enable retrieval of essential information characterizing the potential field, leading to fast acquisition of images with chemical contrast. The control scheme is based on tip sample distance modulation at an angular frequency of ${\omega}$, and null control of the ${1\omega}$ component of the self excitation frequency of the oscillator. It is demonstrated that the control is robust in UHV for a frequency well as small as a few Hz/MHz, and that the mathematical treatment results in satisfactory identification of the potential field. Morse potential is chosen as a case study for identifying the Morse parameters per pixel. Atomic features with similar topography were distinguished by differences in the parameters. The decay length parameter was resolved with a resolution of 10 pm. The method was demonstrated on quenched Silicon at a scan rate comparable to normal imaging.
%\textit{(186 words)}
\end{abstract}

\pacs{37,68,73,82}
\maketitle

In dynamic mode Atomic Force Microscopy, an oscillator, such as a cantilever or a tuning fork is used to vibrate the probing tip in proximity of the sample surface. There are basically two types of operational modes of imaging, which are: (i) the constant frequency shift mode \cite{Giessibi1995}, and (ii) the constant height mode \cite{Guillermet2011, Majzik2016}. The former, although widely used for imaging, risks loosing a large portion of information, since the obtained equi-frequency shift surface is dependent on the imaging parameters and do not necessarily retain information of the potential landscape. The latter has the possibility of yielding more information, since the logged frequency shift can be used to calculate the force or potential \cite{Gross2009, Gross2010}. The mode, ideal as it may sound, can only be used in low drift conditions, due to the absence of tip sample distance regulation, and is often time consuming. In the case of CO terminated tips, it also has the difficulty in probing the surroundings of atomic or molecular protrusions due to masking effects and possible tip tilting mechanisms \cite{VanDerLit2016, Hapala2016}.

With a view to solving the problems listed above, and to implement microscopy with chemical contrast, we introduce here a control scheme that sets the working point to the vicinity of the bottom of the frequency curve (FC), and mathematical treatment to parametrize the acquired data as parameters that define the potential field between the tip and the sample. The control of tip sample distance is accomplished by applying position modulation of the sample in the tip sample direction with an angular frequency of ${\omega}$. Tip sample distance is regulated to null the ${1\omega}$ component of the oscillation. As a result, in the case of low dither amplitudes, the bottom of the FC is maintained as the working point. For larger dither amplitudes, the working point moves away from the sample surface due to asymmetry of the FC.

FC is uniquely defined per site for a given amplitude of drive $A$. Calculation based on the acquired data is facilitated by the boundary conditions assured through the null control of $\Delta f_{1\omega}$. Such position modulation technique was first introduced to SPM as a differential imaging technique \cite{Abraham1988, Sugimoto2009} then applied to lateral position control \cite{Pohl1988, Kawakatsu1990, Kawakatsu1991} or tip sample distance regulation \cite{Kawakatsu2011, Rode2014}.  Tip sample distance modulation was also used for synchronous detection of the harmonics of modulated frequency shifts in order to reconstruct full frequency shift curves between the tip and the sample \cite{Kawai2012}. Here, position modulation of the tip sample distance is set to a few angstroms with a view to scanning, in the vertical direction, the skin depth of the short range forces.

Ideally, not to loose any information, all the higher harmonics of the oscillation should be logged and used for the characterization of the surface. In actual imaging, depending on the amplitude of drive of the fundamental mode of oscillation, the higher harmonics may have much smaller amplitudes, rendering measurement with reasonable frequency noise difficult. Here as a case study, we adopt the Morse potential, and identify the Morse parameters for every pixel of imaging by logging the DC and the  ${2\omega}$ component of the modulated oscillation. Suitability of working with a limited order of higher harmonics needs to be evaluated through the fitting ability of the identified parameters to actual measurement, and through imaging of various samples.

From its easy physical grasp, with its three fitting pa- rameters, the Morse potential is a widely used model to describe inter-atomic potentials  ~\cite{Morse1929, Flugge}, gases and surfaces ~\cite{atkins2006, Girifalco1959, Doye1997, Li2010}, and also widely used for inter-atomic interactions in molecular calculations ~\cite{Person1963, Mahlanen2005, Zhou2002, Tanimura1998, Erkol2009, Chen2002, Vela2005}. The Morse potential is given by:

\begin{equation}
U(Z)= -E_b\left(2 e^{-\frac{Z-Z_0}{\lambda }}-e^{-\frac{2 (Z-Z_0)}{\lambda}}\right)
\label{eq:one}
\end{equation}

Where $E_b$ is the energy depth of the bond, $Z_0$ is the equilibrium distance, and $\lambda$ is the decay length that parametrizes the width of the well. These three parameters serve as chemical fingerprints of the atoms on the surface. 

The frequency shift of a cantilever that oscillates with amplitude $A$ is given by ~\cite{Giessibl1997,Sader2004,Sader2004a}:

\begin{equation}
\resizebox{\columnwidth}{!}{$
\Delta f (z) = \frac{f_0}{k} \frac{2 E_b}{A \lambda } \left(  e^{-2 \frac{Z -Z_0}{\lambda}} e^{-\frac{2A}{\lambda}}
   I_1\left(\frac{2 A}{\lambda}\right)-
e^{- \frac{Z -Z_0}{\lambda}} e^{-\frac{A}{\lambda}}
   I_1\left(\frac{A}{\lambda}\right)\right)
$}
\label{eq:two}
\end{equation}

$f_0$ and $k$ are eigenfrequency and the stiffness of the flexural mode. Modified Bessel functions of the first kind $I_{n}(x)$ are used in preference to Kummer functions for ease of manipulation of the equations. For the specific case where $A<<\lambda$, this frequency shift is proportional to the second derivative of the potential ~\cite{Giessibl1997}. In the general case of amplitude of $A$, if a dither amplitude $D$ is applied around the position $Z_{D0}$ at angular frequency $\omega$ to keep the condition $\Delta f_{1\omega}=0$, we obtain:

\begin{equation}
\Delta f_{1\omega} =\frac{\omega }{\pi } \int_{0}^{\frac{2\pi}{\omega}}\Delta f\left( Z_{D0}+Dsin(\omega t) \right)sin(\omega t) dt = 0
\label{eq:three} 
\end{equation}

Solution of Eq. (\ref{eq:three}) is solved for the measured center of dither $Z_{D0} $
 \begin{equation}
Z_{D0}=Z_{0}+\lambda \times (h_1(\alpha)+h_2(\beta)) 
 \label{eq:four}
\end{equation}
With
\begin{subequations}\label{five}
\begin{align}
h_1(\alpha)=ln\left(\frac{I_1(2\alpha)}{I_1(\alpha)} \right)-\alpha \label{fivea}\\
h_2(\beta)=ln\left(\frac{I_1(2\beta)}{I_1(\beta)} \right) \label{fiveb}
\end{align}
\end{subequations}

$h1(\alpha)$ and $h2(\beta)$ (with $\alpha = A/\lambda$ and $\beta = D/\lambda$) are functions of $A$ and $D$ and therefore dimensionless functions indicating the relative position of $Z_{D0}$ with respect to $Z_0$. $Z_{D0}$ is the controlled value of the tip-sample distance, and can be seen as a universal topography for it is expressed uniquely as a function of the Morse parameters. 

Since $\Delta f$ is a periodic signal around the working point, the $DC$ and $2\omega$ components of $\Delta f$ are used to derive the Morse parameters:

\begin{subequations}\label{six}
\begin{align}
\Delta f_{DC} =\frac{\omega }{2\pi } \int_{0}^{\frac{2\pi}{\omega}}\Delta f\left( Z_{D0}+Dsin(\omega t) \right)dt\\
=\frac{f_0}{k} \frac{E_b}{\lambda^{2}} f(\alpha) g_1(\beta) \label{sixa}\\
\Delta f_{2\omega} =\frac{\omega }{\pi } \int_{0}^{\frac{2\pi}{\omega}}\Delta f\left( Z_{D0}+Dsin(\omega t) \right)cos(2\omega t) dt\\
 =\frac{f_0}{k} \frac{E_b}{\lambda^{2}} f(\alpha) g_2(\beta) \label{sixb}
\end{align}
\end{subequations}
With	
\begin{subequations}\label{seven}
\begin{align}
f(\alpha)=\frac{I_1(\alpha)^2}{2 \alpha I_1(2\alpha)}  \label{sevena}\\
g_1(\beta)=\frac{4  I_1(\beta) \left( I_0(2 \beta) I_1( \beta)- I_0( \beta) I_1(2 \beta)  \right)  }{   I_1(2\beta)^2  }  \label{sevenb}\\
g_2(\beta)=\frac{8  I_1(\beta) \left( I_1(2 \beta) I_2( \beta)- I_1( \beta) I_2(2 \beta)  \right)  }{   I_1(2\beta)^2  }  \label{sevenc}
\end{align}
\end{subequations}

$f$, $g_1$ and $g_2$ are dimensionless functions of $\alpha$ and $\beta$. In the range where $\Delta f_{DC} $ and $\Delta f_{2\omega}$ are monotonic, (i.e. for every $\alpha$ and for $\beta<\approx 2$), the system can be uniquely solved for $E_b$ and $\lambda$. $\lambda$ is injected into Eq. (\ref{eq:four}) to obtain $Z_0$. There are no general analytical expressions for the solution of the system but it can be solved numerically \cite{SupMatALLAIN2016}. 
For $A<<1$ we have : 
\begin{equation}
\Delta f _{min}=-  \frac{1 }{8} \frac{f_0 }{k} \frac{E_b}{\lambda^2}
\end{equation}
and,
\begin{equation}
F_{min}= \frac{E_b}{\lambda} 
\end{equation}

giving a rough idea on the physical meaning of local minima of frequency and force.

\begin{figure}
\includegraphics[width=\linewidth]{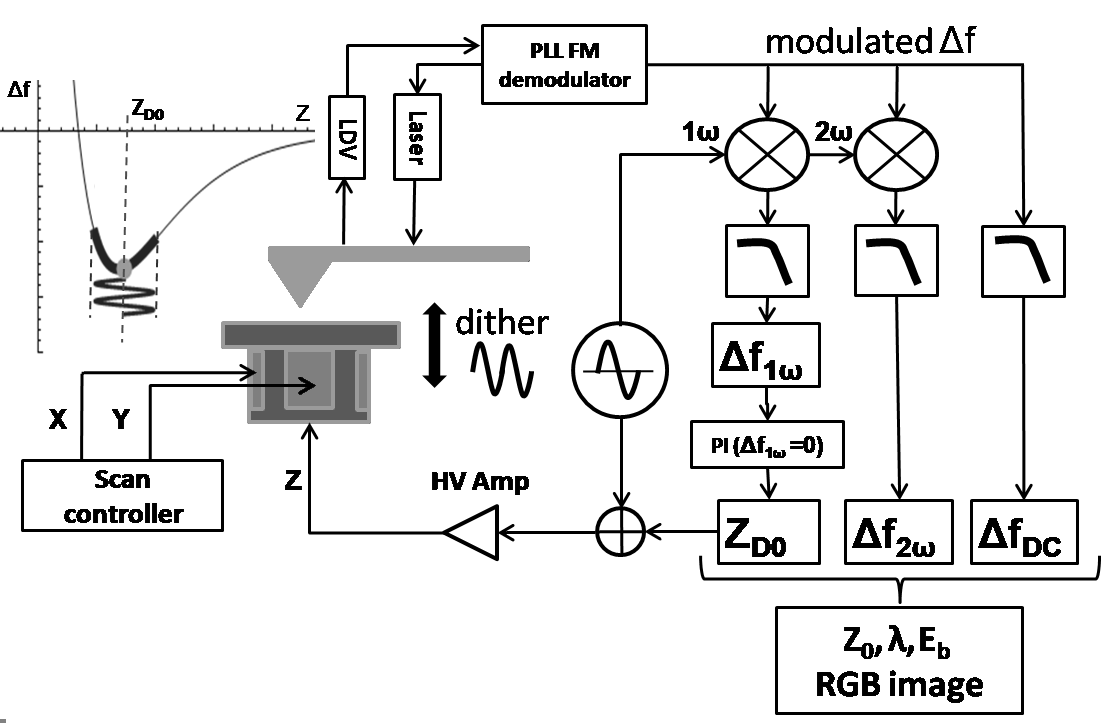}
\caption{\label{fig:blockdiagram} Schematic of the parameter mapping method. LDV: Laser Doppler Vibrometer. PI: Proportional Integral Feedback controller.}
\end{figure} 

Figure \ref{fig:blockdiagram} depicts the schematic of the AFM. The $\Delta f_{1\omega}$ signal is fed to the z feedback control loop to keep its value close to zero. Angular frequency ${\omega}$ is set below the frequency of self-excitation around 2 MHz and above the cut-off frequency of the feedback control.

\begin{figure}
\includegraphics[width=\linewidth]{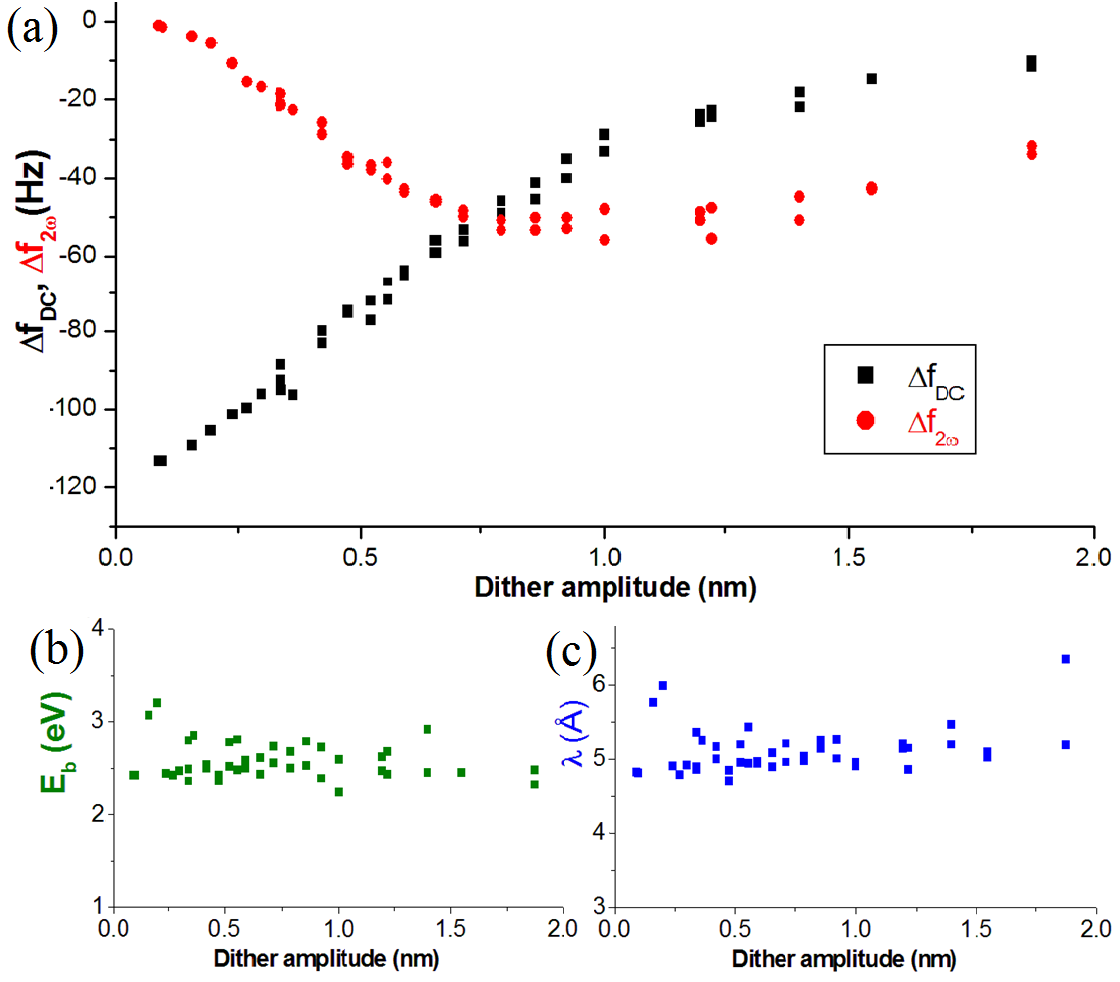}
\caption{\label{fig:experimental1} Determination of $E_b$ and $\lambda$ versus dither amplitude on top of Si(111) from experimentally obtained frequency components $\Delta f_{DC}$, $\Delta f_{2\omega}$ and solving equations \eqref{sixa} and \eqref{sixb}. (a) Experimental values of $\Delta f_{DC}$ and $\Delta f_{2\omega}$. (b) Effective bonding energy $E_b$. (c) Effective decay length $\lambda$.  Acquisition parameters were $f_{2nd}\approx 1.99$ MHz, $k_{2nd} = 2,400~N.m^{-1}$, $A$ = 7 \AA, $\omega$ = 1 kHz. }
\end{figure} 

Figure \ref{fig:experimental1} shows dependence of calculated $E_b$ and $\lambda$ on $D$ while executing null control of $\Delta f_{1\omega}$. The values are more or less insensitive to $D$, which is a good sign, since the identified Morse parameters should be intrinsic to the site and not affected by experimental conditions such as $D$. Small $D$ should be avoided since the tip apex does not make sufficient excursion within the short range force, making the  ${2\omega}$ signal too weak. Large $D$ should also be avoided since $\Delta f_{DC}$ converges to zero due to excessive averaging of the FC well. 

Figure \ref{fig:experimental1}(a) depicts the measured values of $\Delta f_{DC}$ and $\Delta f_{2\omega}$. In simulation and in actual measurement, signals $\Delta f_{DC}$, $\Delta f_{2\omega}$ cross around  $0.75 \lambda$, which gives a guideline on the favorable value of $D$ for minimizing the measurement noise. The values in Figs.\ref{fig:experimental1}(b) and \ref{fig:experimental1}(c) are snap shots of the calculated Morse parameters. The values appear to be scattered for amplitudes of $D$ far from the cross-over point but otherwise give compatible noise level for scanning experiments. Imaging shows that the values are site dependent and sufficiently low-noise to differentiate the atomic features.

\begin{figure*}
\includegraphics[scale=0.58]{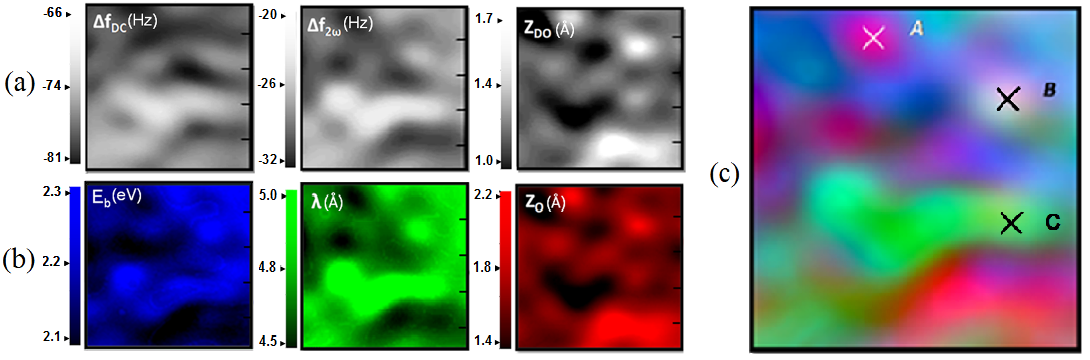}
\caption{\label{fig:scheme}  Morse parameters mapping on quenched Si(111). (a) Measured quantities using Eqs.(\ref{eq:four}), (\ref{sixa}) and (\ref{sixb}). (b) Calculated Morse parameters. (c) RGB Morse parameters mapping. Image size is 2.2 x 2.2 $nm^{2}$. Acquisition parameters were $f_{2nd}\approx 1.99$ MHz, $k_{2nd} = 2,400~N.m^{-1}$, A = 7 \AA , $\omega$ = 1 kHz and $D$ = 4 \AA.}
\end{figure*}
\begin{figure*}
\includegraphics[scale=0.62]{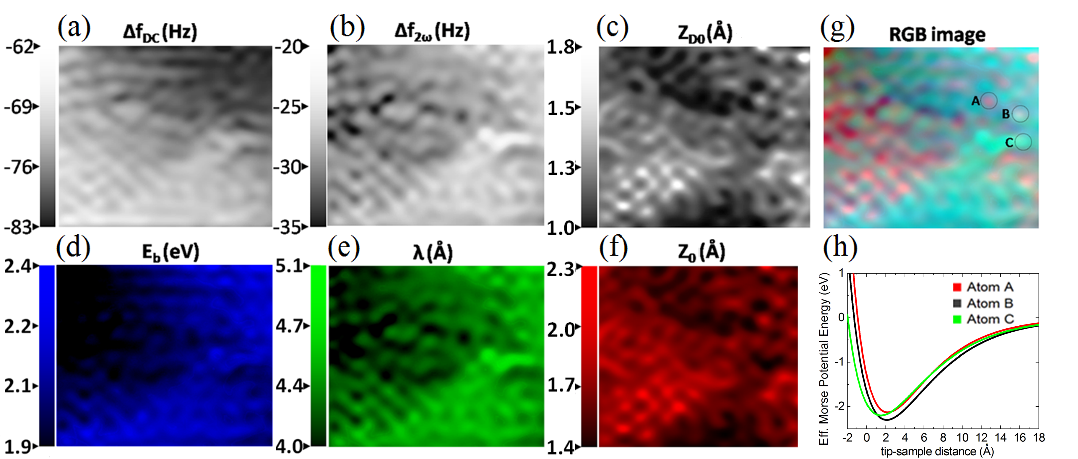}
\caption{\label{fig:experimental2} Experimental parameter mapping results on quenched Si(111). The dimension of each picture is 6x5 $nm^{2}$. Scale units are stated above the scale bars. Measured values : (a) $\Delta f_{DC}$, (b) $\Delta f_{2\omega}$ and (c) $Z_{D0}$. Calculated Effective Morse Parameters : (d) $E_b$, (e) $\lambda$, (f) $Z_0$ and (g) RGB combined picture (R = $Z_0$; G = $\lambda$; B = $E_b$) with distinguishable atoms. (h) reconstructed Morse potential profile for atoms A, B and C. Acquisition parameters were $f_{2nd}$ = 1.991 961 MHz, $k_{2nd} = 2,400~N.m^{-1}$, $A$ = 7 \AA, $\omega$ = 1 kHz and $D$= 4 \AA.}
\end{figure*} 

Figures \ref{fig:scheme} and \ref{fig:experimental2} depict signals $\Delta f_{DC}$, $\Delta f_{2\omega}$ and $Z_{D0}$ acquired on quenched Si(111) while keeping the ${1\omega}$ component of $\Delta f$ to zero. $E_b$, $\lambda$ and $Z_{0}$ were numerically calculated from the acquired data. Details on the AFM can be found in the literature \cite{Umeda1991, Nishida2009, Hoel2012}.

\begin{figure}
\includegraphics[scale=0.2]{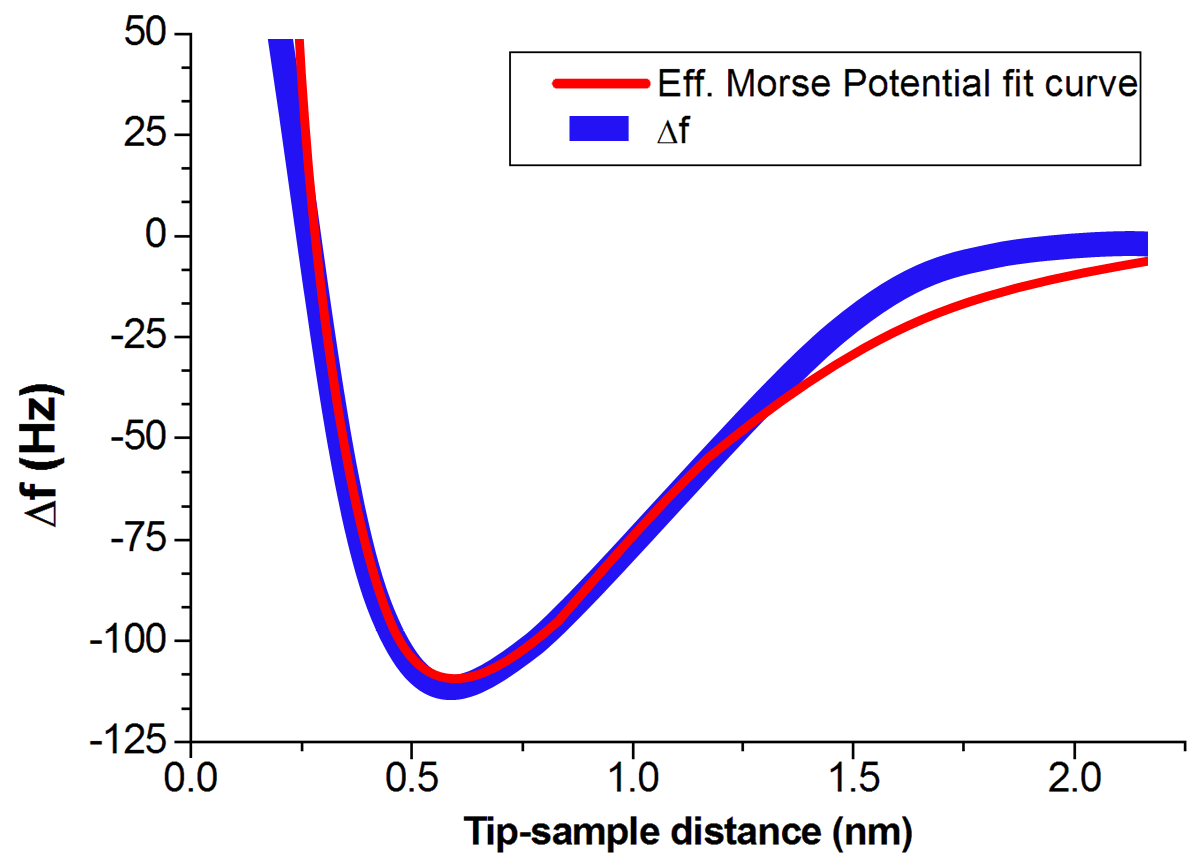}
\caption{\label{fig:spectroscopy1} Curve fitting between measured FC and FC generated from effective Morse Parameters,  $E_b = 2.2~eV$ and $\lambda = 4.6$ \AA$ $ and using Eq. \eqref{eq:two}. Same paremeters were used as Fig. \ref{fig:experimental1} }
\end{figure} 

Figure \ref{fig:spectroscopy1} depicts a measured FC and a FC generated from a set of Morse Parameters chosen to fit the measurement. The fit is satisfactory around the bottom of the FC where the working point is modulated by $D$, but deviates further away from the sample surface. This may be due to the commonly known fact that the Morse Potential is better adapted for smaller inter-atomic distances \cite{Flugge}. Adopting a potential field model with more parameters may result in a better curve fit of the FC. What we propose is to apply the method to various samples to begin with, and if it proves to be limited, move onto other potential models using measured higher harmonics. One may note that the fit is already satisfactory within the range of the position modulation $D$, and that larger $D$ may be needed to identify parameters that fit the entire FC.

From the experimental results, we have demonstrated the following:
(i) atomic resolution is obtained in UHV for the proposed control method, and control is stable even for a small local minima of frequency of 5 Hz over 2 MHz.
(ii) Calculation from the measured data to the Morse parameters took 1 sec per pixel, but could be shortened to 2 ms by preparing a look-up table for the conversion,
(iii) values of $E_{b}$ and $\lambda$ tend to be larger than expected for a Si-Si pairwise potential,
(iv) if atomic features take the same colour, it means that the set of Morse parameters are the same. However, care is needed in interpretation of the images. Atomic features  A and B are similar atomic features in the topography $Z_{D0}$ image, but different in the final colour image due to difference of $\lambda$. Difference in $\lambda$ could be resolved in the 10 pm order. Features B and C have more or less the same $\lambda$ and $E_{b}$ but differ in colour due to the topographic difference. The derived Morse parameter  $Z_{0}$ is directly affected by topography such as atomic steps. The sample should be atomically flat, or the topography channel (Red) should be used with care.
The method proposed here starts off with measured FC and derives a set of Morse parameters that give a satisfactory curve fit between measured and calculated FC. The measured FC is a result of various interaction between the tip and the sample, so due to such effects as back ground forces and relaxation, the derived Morse parameters may deviate from the pairwise potential expected of the tip apex atom and the sample atom. Deviation needs to be assessed in a quantitative manner to deduce Morse parameters close to the pairwise potential. If we limit the discussion to an atomically flat surface, the discussion can be simplified to the  following: (i) tip volume acts to increase van der Waals force, resulting in larger $E_{b}$ and larger relaxation of the tip and the sample. The latter results in a larger $\lambda$ due to fishing-out and pushing-in of the sample atom. Through calculation, placing a molecule at the apex of an AFM tip has the following effects in view of the method proposed: (i) Due to van der Waals attraction, effective $E_{b}$ is minimal when tip radius R is zero, and increases with R. Effective $E_{b}$ is 2 times larger for R$\approx$10 nm \cite{SupMatALLAIN2016}. Functionalising the tip apex with a molecule \cite{Guggisberg2000} greatly lessens the increase of effective $E_{b}$ due to R, (ii) for a rigid body model $\lambda$ is much less sensitive to increase of R. However, if relaxation of the sample and the tip are taken into consideration, a sharper tip or a molecular functionalized tip is effective in lessening the difference between effective $\lambda$ and that of the pairwise potential. As a rough calculation, assuming the spring constant of the sample atom to its surrounding to be 10 N/m, attractive force of 160 pN results in 16 pm of relaxation that is not accounted for in calculating $\lambda$. Calibration of the same order is necessary in the repulsive regime. It was demonstrated that the method allows resolving the difference in effective $\lambda$ in the order of 10 pm. This difference can be ascribed to at least two effects, namely, (i) the difference of the pairwise potential, and (ii) the difference of bonding rigidity of the sample atom to its surroundings. Further study and simulation is needed to access the ratio of the two contributions to the contrast mechanism. As mentioned above, introducing a molecular terminated tip, such as an adamantine molecule terminated tip \cite{Katano2007}, will act in favour in lessening the difference between effective $E_{b}$, $\lambda$ and the pairwise $E_{b}$ and $\lambda$. Using a well defined molecular tip will improve the repeatability of the measurements, and quantitative evaluation of the sample.

The proposed method is robust, and in many cases, tip change does not occur during acquisition of an image. This makes relative comparison of the atomic features within the image possible. However, to establish quantitative observation of the surface, a reference tip termination method needs to be sought for, and the derived physical values studied in terms of resolution and absolute values. Effects of combination of neighbouring atoms and subsurface compositions need also be studied, and a database constructed for a better understanding. The relatively simple and robust method  introduced here offers such possibilities. We hope that the method will serve as a milestone towards such efforts. 

The authors would like to thank the CNRS, IIS-UTokyo, the JSPS Postdoctoral Fellowship and Core-to-Core Programs for financial support. We thank Markku Kainlauri, Frank Rose, Anthony Genot and Elsa Mazari Arrighi for their scientific feedback. H. K. thanks Norio Miyata for valuable advice at the early stages of the project.

\bibliography{ColorAFM}

\end{document}